# Effect of ZrH$_2$ particles on the microstructure and mechanical properties of IN718 manufactured by selective laser melting

C. Galera-Rueda[1,2], E. Nieto-Valeiras[1,2], M. Gardon[3],

M.T. Pérez-Prado[1], J. LLorca[1,2,*]

[1] IMDEA Materials Institute, Calle Eric Kandel, 2, 28906 Getafe, Madrid, Spain.

[2] Department of Materials Science, Polytechnic University of Madrid/Universidad Politécnica de Madrid, E. T. S. de Ingenieros de Caminos, Madrid 28040, Spain.

[3] Renishaw Ibérica, De de la Recerca, 7, 08850 Gavá, Barcelona, Spain

*Corresponding author

**Abstract**

The influence of Zr additions (in the form of ZrH$_2$ particles) on the microstructure and mechanical properties of IN718 Ni-based superalloy manufactured by selective laser melting was explored. Fully dense alloys could be obtained by careful selection of the processing parameters. The addition of ZrH$_2$ did not modify the microstructure of the alloy but introduced a dispersion of Zr-rich globular particles of approximately 50 nm in diameter which increased the strength of the as-printed material at ambient (23ºC) and high temperature (550ºC). Heat treatments of solubilisation followed by aging led to a fine dispersion of $\gamma''$ precipitates that controlled the strength of the alloy, which was independent of the addition of Zr. Moreover, dynamic strain ageing was found in the heat treated materials when deformed at 550ºC. Finally, the strength of the IN718 deformed perpendicular to the building direction was higher than that along the building direction, regardless of the heat treatment and temperature. Computational homogenization simulations showed that the anisotropy was associated with the strong texture.

# 1. Introduction

Nickel-based superalloys are widely used in high temperature structural applications because of their outstanding mechanical properties and corrosion resistance. Inconel 718 (IN718), developed by International Nickel Company in the late 1950's, accounts for approximately one-third of all superalloy production [1] and polycrystalline IN718 is extensively used for intermediate temperature applications (up to 600-700 ºC) in gas turbines and aeroengines [2]. The microstructure of IN718 is made up by a $\gamma$ phase (Ni FCC solid solution) which contains a dispersion of nm-sized $\gamma'$-$Ni_3$(Al,Ti) and $\gamma''$-$Ni_3Nb$ coherent precipitates within the grains together with μm-sized metal carbides and $\delta$ phase ($Ni_3Nb$) particles at grain boundaries [3-5]. This alloy is produced in both cast and wrought forms and has a great processing versatility, e. g. the microstructure and properties can be tailored by tuning the fabrication temperatures, forging reductions, and strain rates [1].

More recently, IN718 has started to be used to manufacture components with complex geometries by means of selective laser melting (SLM) because of its good weldability and reduced hot cracking tendency [6-7]. While the porosity in IN718 processed by SLM can be reduced to a minimum with a proper combination of parameters [8], the mechanical properties are anisotropic due to the presence of elongated grains and a marked (100) crystallographic texture in the building direction [9-11]. In general, the strength and elongation to failure of IN718 processed by SLM in the building direction are similar those found in the wrought alloy. However, the strength increases and the elongation to failure decreases in the orientation perpendicular to the building direction and these trends remain after heat treatments [6-7].

Moreover, microstructural inhomogeneities in the form of Nb segregation as well as residual stresses are also associated with IN718 processed by SLM [10,12] and different strategies have been used to overcome these limitations. The first approach to improve the microstructure and properties of SLM-processed IN718 is to optimize the processing parameters (laser power and speed, layer thickness, pre-heating temperature, and hatch distance). This strategy has a very important effect in the porosity levels but it was inefficient to provide a change in texture because of the dominant role of the thermal gradient, which is inherent of the printing process [9]. The use of a rotational scanning strategy has been able to reduce the intensity of the texture [13-15] but the anisotropy in the properties was still too large.

Heat treatments are another option to improve the properties of IN718 manufactured by SLM. Standard solution and precipitation heat treatments improve the strength by the precipitation of γ' and γ'' but the ductility is impaired by the precipitation of acicular δ phase [16]. This problem can be overcome if homogenization treatments in combination with hot-isostatic pressing are applied before the solution and precipitation treatments [17-18] but this strategy increases substantially the processing costs.

Finally, other investigations modified the microstructure and properties of IN718 manufactured by SLM through the addition of carbide particles [19-21]. Addition of TiC led to a reduction in segregation of Nb because the formation of NbC was favoured [19] and to an increase in strength and wear resistance [19-20] although no information of the ductility was provided. It is known, however, that the addition of TiC reduced the ductility of IN625 manufactured by SLM [22].

Previous investigations in cast IN718 have shown that addition of small amounts of Zr led to a refinement of the grain size, improving the hot cracking resistance and the machinability of the alloy [23]. In parallel, the addition of Zr to cast CMC247LC superalloy did not reduce the grain size but changed the shape of the carbides from angular to globular [24] and this modification improve the strength and the creep resistance. Nevertheless, this strategy has not been explored in the case of IN718 manufactured by SLM and this is the objective of this investigation. To this end, following the approach developed in [25], Zr additions in the form of $ZrH_2$ particles were mixed with the IN718 powders. Dog-bone tensile specimens were manufactured by SLM to understand the effect of Zr on the microstructure and mechanical properties at ambient and elevated temperature.

## 2. Experimental techniques

Gas-atomized IN718 powders were supplied by Renishaw while angular-shaped $ZrH_2$ powders were obtained from Hongwu International. The particle size distributions were measured by laser diffraction in an ethanol medium (Mastersizer 2000, Malvern). IN718 was mixed with 1% wt. $ZrH_2$ in a planetary mill (Pulverisette 6) without any grinding medium in cycles of 5 min at 200 rpm followed by a pause of 10 minutes during 24 hours.

SLM was carried out in a AM400 machine (Renishaw) using Reduced Build Volume accessory, which permits operating with a low amount of powder (around 0.75 L). Bidirectional scanning strategy with 67° rotation per layer was selected for both alloys. Laser speed was fixed at 0.875 m/s and layer thickness at 30 µm while the laser power (P) and the hatch distance (H) were varied as indicated in Table 1, leading to a range of energy densities from 42 to 127 J/mm$^3$.

Table 1. Parameters of SLM for IN718 and IN718 + ZrH$_2$ and densities.

|   | Power (W) | Hatch distance (µm) | Energy density (J/mm$^3$) | Density IN718 (%) | Density In718+ZrH$_2$ (%) |
|---|---|---|---|---|---|
| 1 | 100 | 90 | 42 | 96.38 | 96.13 |
| **2** | **200** | **90** | **85** | **99.96** | **99.85** |
| 3 | 200 | 80 | 95 | 99.91 | 99.72 |
| 4 | 250 | 90 | 106 | 99.89 | 99.70 |
| 5 | 300 | 90 | 127 | 99.68 | 99.70 |
| 6 | 300 | 80 | 143 | 99.43 | 99.27 |

Specimens with the shape of a trunked pyramid attached to a 10 x 10 x 5 mm$^3$ parallelepiped were manufactured using these conditions. After processing, the parallelepipeds were cut parallel and perpendicular to the building direction using a disc cutting machine with a resin-embedded SiC saw. Samples were grinded using SiC grinding paper from 320 to 2000 grit and polished using cloths with diamond pastes of 6, 3 and 1 µm. The porosity in the cross-sections was measured using ImageJ from optical microscopy images obtained with an Olympus BX-51 microscope. Heat treatments were performed in the specimens with the highest density in each material in a muffle furnace. They involve a solubilisation step at 982°C for 1 hour, water quenching and a two-step aging treatment which included 8 hours at 718°C followed by 10 hours at 621°C. The specimens were finally air cooled.

The microstructural characterization was carried out in the samples with the highest density. They were examined by electron backscattered diffraction (EBSD) using a field-emission gun scanning electron microscope (SEM) (Helios NanoLab 600i, FEI) equipped with an HKL EBSD system, a CCD camera, as well as Aztec and Channel 5.0 data acquisition and analysis software packages, respectively. EBSD measurements were conducted using a step size of 0.8 µm, an accelerating voltage of 20 kV, and a beam current of 2.7 nA. Texture was measured with the EBSD data using Channel 5.0. Composition was measured by Energy Dispersive X-rays spectroscopy (EDS) in the same microscope with a Oxford Instruments X-Max$^N$ detector. Sample preparation for EBSD and EDS included grinding and polishing with cloths and diamond paste up to 1 µm and a final polishing step with a colloidal silica suspension.

The macrotexture was measured by X-ray diffraction (XRD) in a Malvern Panalytical diffractometer furnished with a five-axis cradle with parallel and filtered Cu Kα radiation (45 kV, 40 mA). Rotations in the 3 angles $\varphi$, $\theta$ and $\psi$ were performed to ascertain the intensity of the peaks corresponding to the planes (200), (110), (111) and (311). The orientation distribution function was obtained using X'Pert Texture software.

Specimens for transmission electron microscopy (TEM) were cut from the printed parallelepipeds using a wire cutting machine. They were manually polished down to ~80 µm and discs of 3 mm in diameter were mechanically punched. The thinned discs were twin jet electropolished at -25°C and 20V in an electrolyte made up of 10% perchloric acid, 30% butanol and 60% ethanol (TenuPol-5, Struers). Bright field (BF) and Z contrast imaging by high angle annular dark field scanning transmission electron microscopy (HAADF-STEM) were performed using a Talos F200X FEI microscope operating at 200 kV. Chemical composition maps were obtained under STEM mode with a SuperX energy-dispersive X-ray (EDX) detector.

Vickers hardness measurements were performed in a HMV-6 Shimadzu microhardness indenter using a load of 1 Kg and a dwell time of 10 s. Five measurements were carried out in each condition and sample. Parallelepipedic samples of dimensions of 33 x 22 x 3.5 mm$^3$ were prepared by SLM using the conditions for minimum porosity with the longest axis parallel to the building direction for both materials and also with the longest axis perpendicular to the building direction in the case of IN718. Dog-bone specimens for tensile tests were prepared from the samples by electro-discharge machining, leading to specimens with a gauge length of 10 mm, 3 mm in width and 2.5 mm in thickness. Tensile tests were performed in a universal mechanical testing machine (Instron 3384) under displacement control at an average strain rate of $10^{-3}$ s$^{-1}$.

# 3. Results

## 3.1. Microstructure

Secondary electron SEM images of the as-received IN718 and ZrH$_2$ powders are shown in Figs. 1a and 1b, respectively. IN718 particles were rounded with D20 = 19.1 µm, D50 = 30.4 µm and D90 = 47.5 µm. ZrH$_2$ particles have irregular shape and D10 = 1.63 µm, D50 = 3.7 µm and D90 = 8.2 µm (Fig. 1d). After mixing, the smaller ZrH$_2$ particles decorated the larger IN718 particles (Fig. 1c).

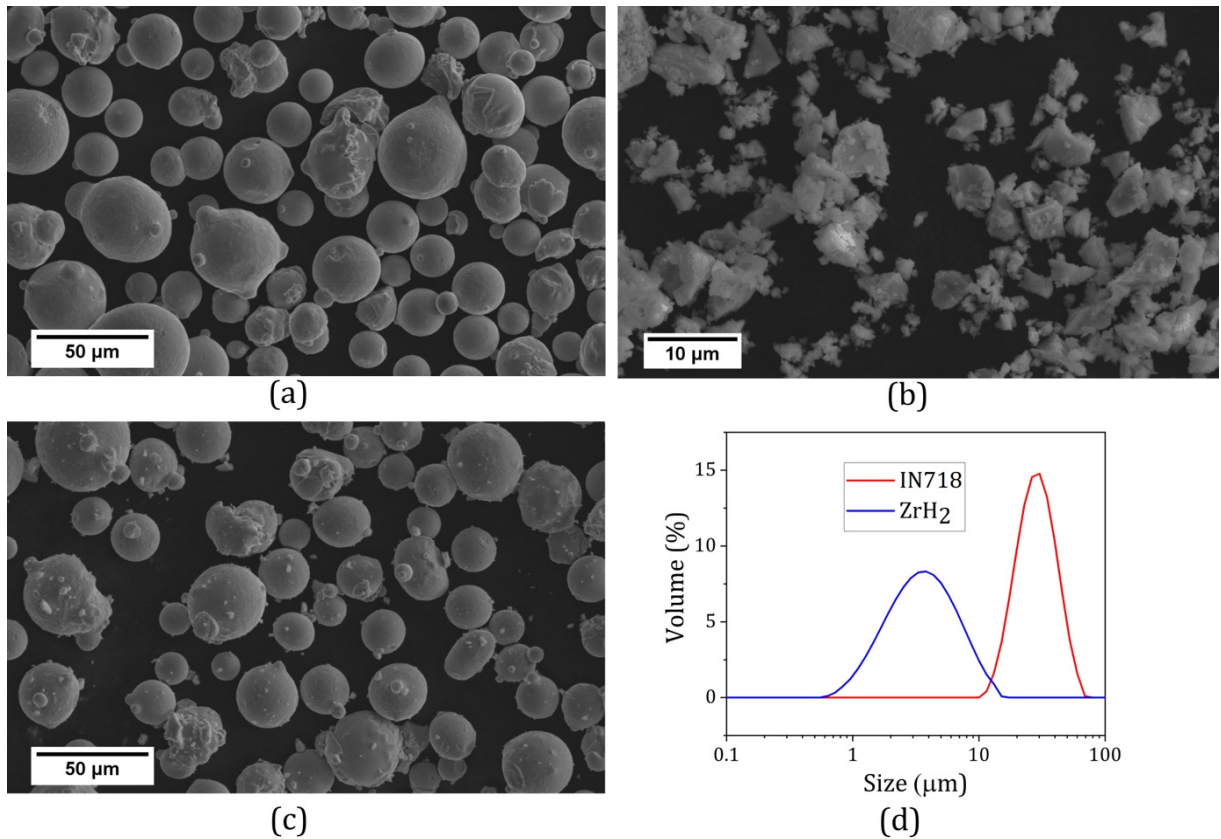

**Fig. 1.** SEM images of the (a) IN718 powder, (b) ZrH$_2$ powder and (c) mixture of IN718 with 1 wt%. of ZrH$_2$. (d) Particle size distributions of IN718 and ZrH$_2$ powders.

The chemical composition of the printed samples is shown in Table 2. The addition of the ZrH$_2$ particles leads to a presence of 0.9 wt. % of Zr in the alloys without any significant changes in the content of other elements. The porosity of the specimens processed by SLM was small in all cases (Table 1) but it was always slightly higher in the IN718+ZrH$_2$ material. Moreover, the addition of Zr led to the appearance of irregular defects. The latter have often been attributed to the decrease of flowability induced by the small particles [26] although sometimes they were also associated with the segregation of Zr, as shown in Fig. 2. Specimens with the highest density of both materials (processed with an energy density of 85 J/mm$^3$) were selected for further investigation.

Al alloys with ZrH$_2$ particle additions processed by SLM showed pores associated with the formation of H$_2$ during the decomposition of ZrH2 [25]. They were not found in the IN718+ZrH$_2$ material because only 1 wt. % of ZrH$_2$ was added (as compared with 3 wt. % in [25]) and the solubility of hydrogen in nickel is higher than in aluminum [27].

Table 2. Chemical composition of the printed alloys (wt. %)

|  | Ni | Cr | Fe | Nb | Mo | Ti | Al | Zr |
|---|---|---|---|---|---|---|---|---|
| **IN718** | 52.53 | 19.67 | 18.14 | 5.28 | 2.81 | 1.02 | 0.54 | - |
| **IN718+ZrH$_2$** | 52.27 | 19.13 | 18.23 | 5.03 | 2.95 | 0.97 | 0.50 | 0.92 |

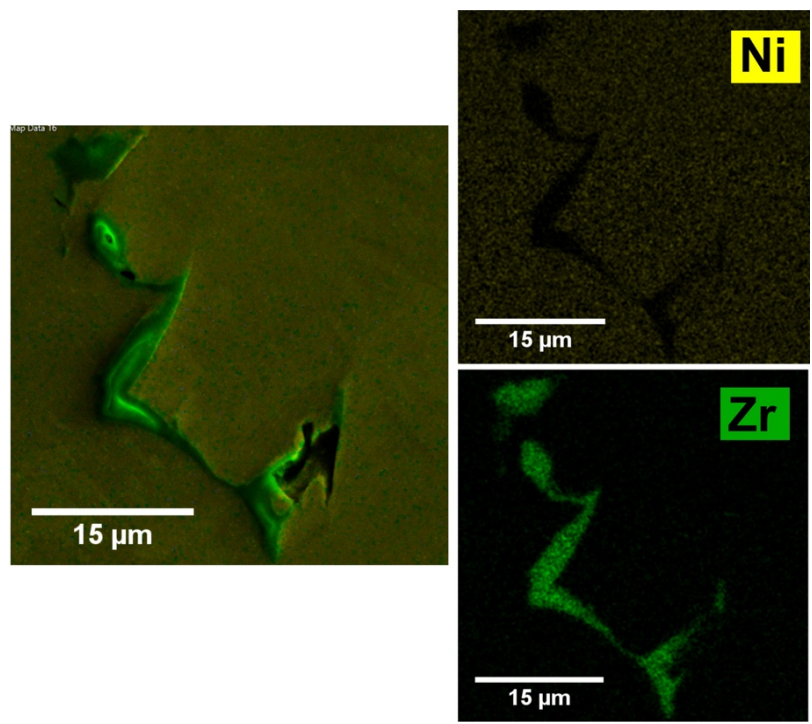

**Fig. 2.** Irregular Zr segregation in the as-printed IN718-ZrH$_2$ material. The EDX maps of Zr and Ni are shown to illustrate the chemical segregation.

The addition of ZrH$_2$ particles did not modify significantly neither the grain size and shape, as shown in Figs. 3 and 4. The cumulative probability plot of the grain area in Fig. 4 indicates that the addition of ZrH$_2$ reduced slightly the grain size of IN718 and that the heat treatment increased -also slightly- the grain size in both alloys (Table 3). Nevertheless, the differences were not relevant from the viewpoint of mechanical properties and did not indicate that Zr particles contribute to grain nucleation.

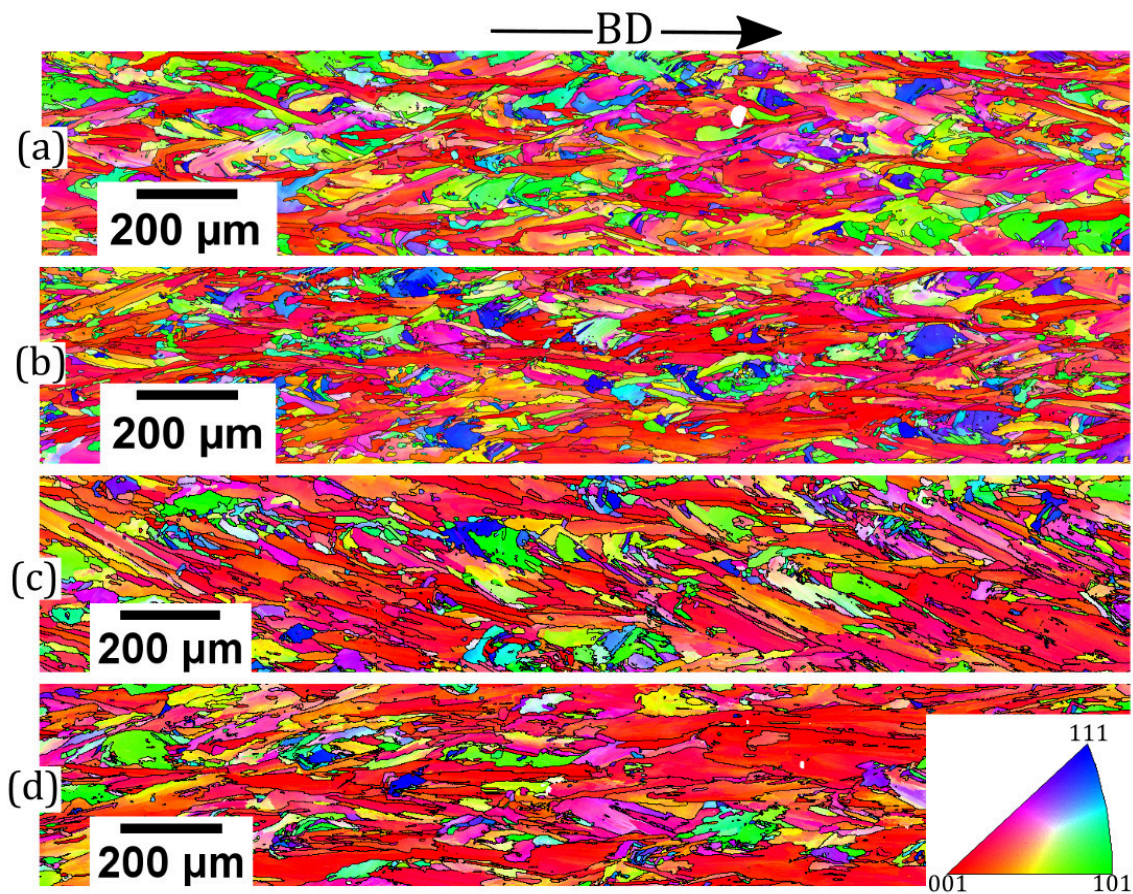

**Fig. 3.** (a) Inverse pole figure EBSD map of the as-printed IN718 alloy. (b) *Idem* of the as-printed IN718+ZrH$_2$ alloy. (c) *Idem* of the heat-treated IN718 and (d) *idem* of the heat-treated IN718+ZrH$_2$ alloy. The building direction is horizontal. High angle grain boundaries (misorientation > 15º) are marked with black lines.

Both alloys exhibit the typical microstructure of IN718 processed by SLM, with elongated grains with the longest axis parallel to the building direction and a strong texture with the (100) plane perpendicular to the building direction [9-11] (Figs 3 and 5 and Table 3). The average grain aspect ratio was in the range 3 to 4, with a very wide distribution, and it was again independent of the addition of ZrH$_2$ particles and of the heat treatment.

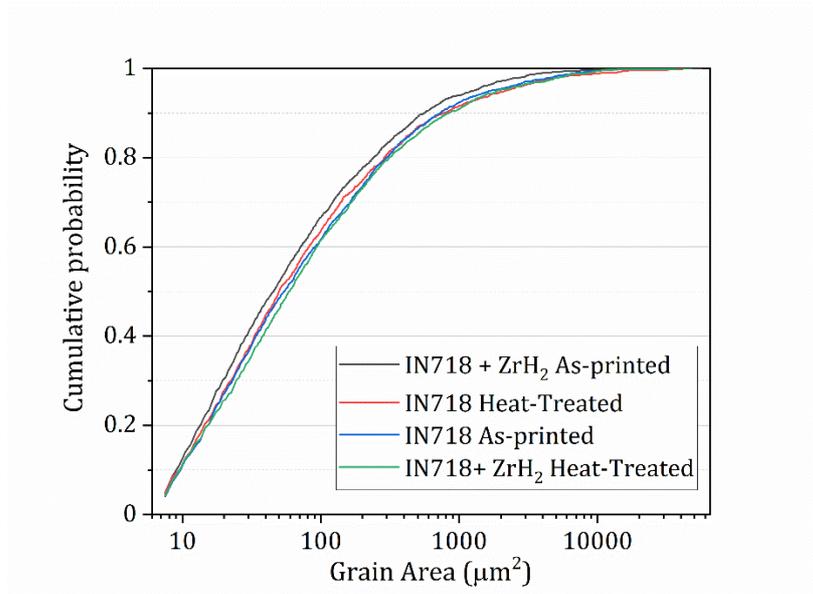

**Fig. 4.** (a) Cumulative probability plot of the grain area of IN718 and IN718+ ZrH$_2$ alloys in the as-printed and heat-treated conditions obtained from EBSD maps of sections parallel to the building direction.

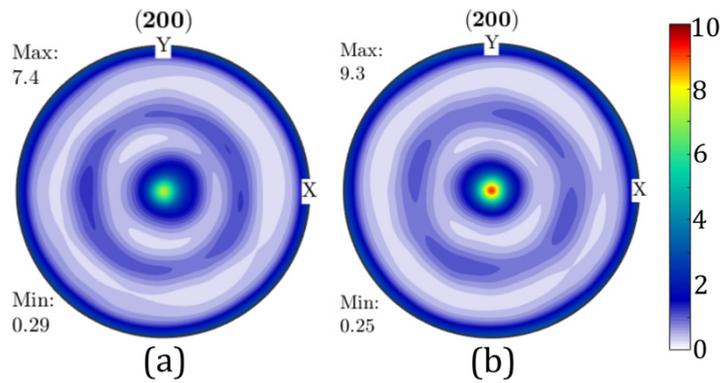

**Fig. 5.** (a) Pole figure of the (200) plane of the as-printed IN718 material. (b) *Idem* of the as-printed IN718-ZrH$_2$ material. X and Y are perpendicular to the building direction. Intensity is measured in mrd.

Table 3. Grain size and aspect ratio obtained from EBSD analysis of sections parallel to the building direction.

|  | Average grain area (μm$^2$) | Aspect ratio | Number of grains |
|---|---|---|---|
| IN718 as-printed | 406 | 3.7 ± 2.5 | 2340 |
| IN718+ZrH$_2$ as-printed | 307 | 3.6 ± 2.5 | 3150 |
| IN718 heat-treated | 522 | 4.4 ± 3.1 | 1502 |
| IN718+ZrH$_2$ heat-treated | 476 | 3.3 ± 2.3 | 2080 |

The microstructure of both alloys in the as-printed and heat treated conditions was analyzed by TEM. The as-printed alloy presented a cellular structure and the cells were elongated along the building direction (Fig. 6). The cell boundaries contained Ni and Ti-rich precipitates, which have been identified as $Ni_3(Nb, Ti)$ δ phase in previous investigations [10, 28]. The cellular structure was maintained in the as-printed alloy containing $ZrH_2$ particles (Fig. 6) and precipitates were found at the cell borders (marked with red arrows in Fig. 7) and within the cells (marked with blue arrows in Fig. 7). The element composition map in Fig. 8 shows some differences with respect to the as-printed IN718. Only Nb (and not Ti) was segregated to the cell boundaries and, in addition, globular Nb-rich precipitates as well as Zr-rich globular particles with a diameter of ≈ 50 nm were dispersed within the cells. The element composition map indicates that these Zr-rich particles may be carbides and/or segregated Zr but not oxides.

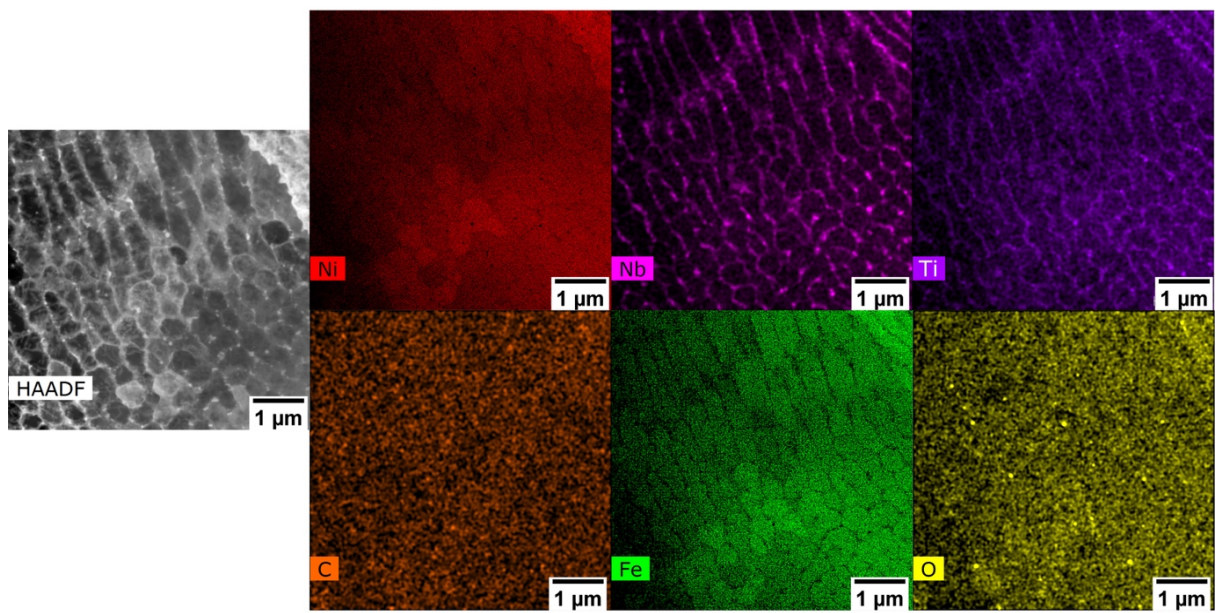

**Fig. 6.** HAADF-STEM micrograph of the as-printed IN718 alloy showing the cellular structure and the corresponding element maps obtained by EDX.

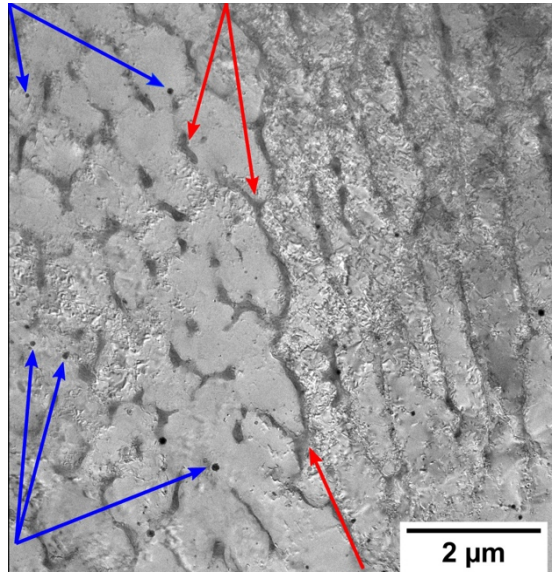

**Fig. 7.** HAADF-STEM micrograph of the as-printed IN718+ZrH$_2$ alloy showing the cellular structure. Precipitates at the cell boundaries are marked with red arrows and within the cells with blue arrows.

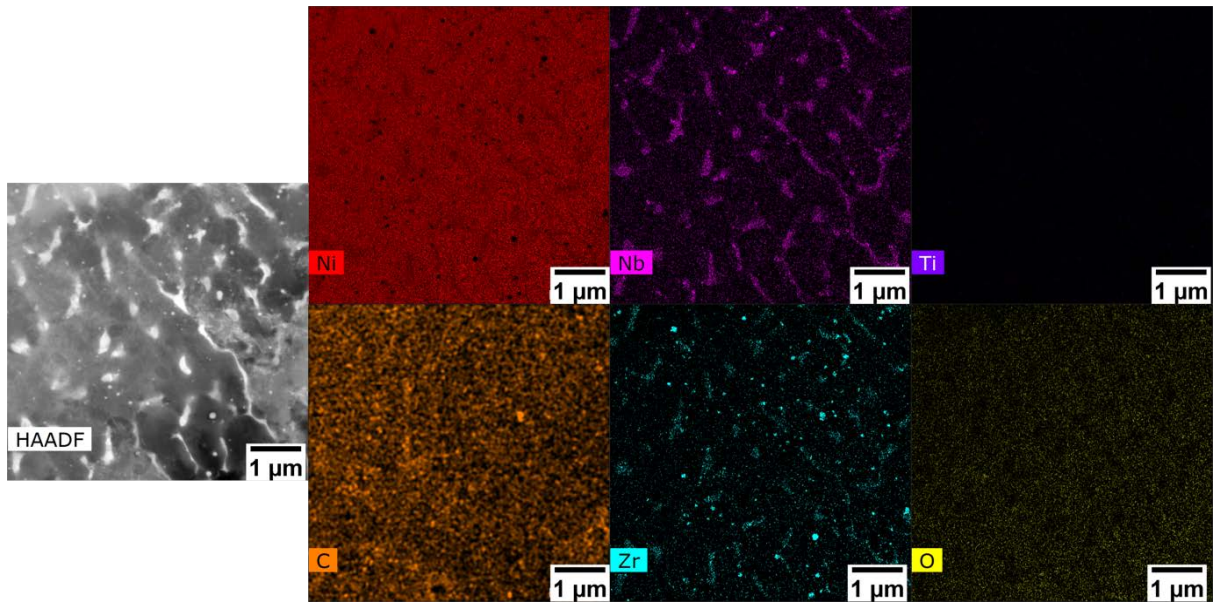

**Fig. 8.** HAADF-STEM micrograph of the as-printed IN718+ZrH$_2$ alloy and the corresponding element maps obtained by EDX.

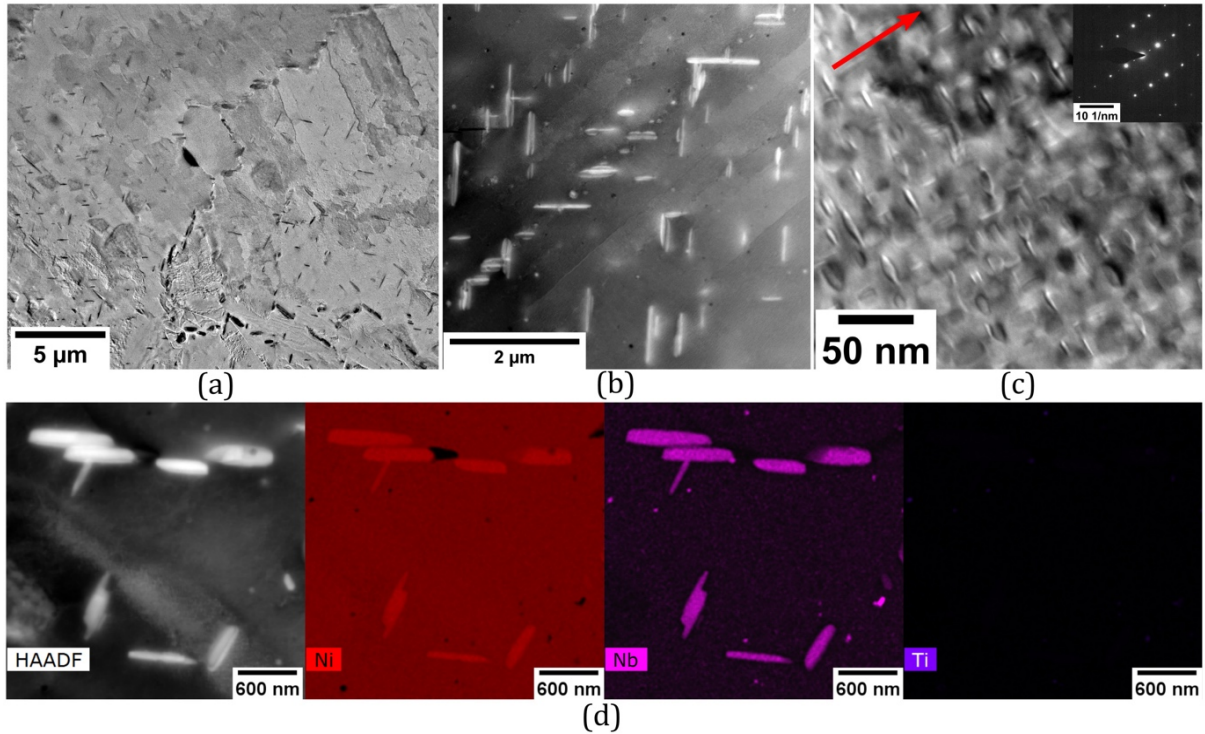

**Fig. 9.** TEM analysis of the heat-treated IN718 alloy. (a) Cellular structure; (b) HAADF-STEM micrograph of acicular δ precipitates at the grain boundaries and within the grains; (c) HAADF-STEM micrograph showing a fine dispersion of disk-shaped γ''-Ni$_3$Nb precipitates, which grew parallel to the (001) planes of the Ni FCC lattice. The zone axis of this micrograph is [013] and the red arrow corresponds to the <200> direction. (d) HAADF-STEM analysis showing the EDX element maps corresponding to the acicular δ precipitates.

The heat-treated IN718 specimens maintained the cell structure (Fig. 9a) but acicular δ precipitates with an average length of the order of 0.5 μm were found at the grain boundaries and within the grains [16] (Fig. 9b). These precipitates contained Ni and Nb but not Ti (Fig. 9d). In addition, the heat treatment led to a fine dispersion of disk-shaped γ''-Ni$_3$Nb precipitates, which grew parallel to the (001) planes of the Ni FCC lattice (Fig. 9c). The addition of Zr particles modified the microstructure after the heat treatment. While the cell structure was maintained (Fig. 10a), the acicular δ precipitates at the cell boundaries were replaced by smaller, rounded δ precipitates (Fig. 10b). In addition, smaller precipitates were found within the cells (Fig. 10b). The element composition map of the δ precipitates showed that they were made up on Ni and Nb (no Ti) while the smaller precipitates within the cells contained Zr, O or C, being more likely Zr carbides and oxides (Fig. 10d). Finally, the fine dispersion of disk-shape γ''-Ni$_3$Nb precipitates was also found at high magnification (Fig. 10c)

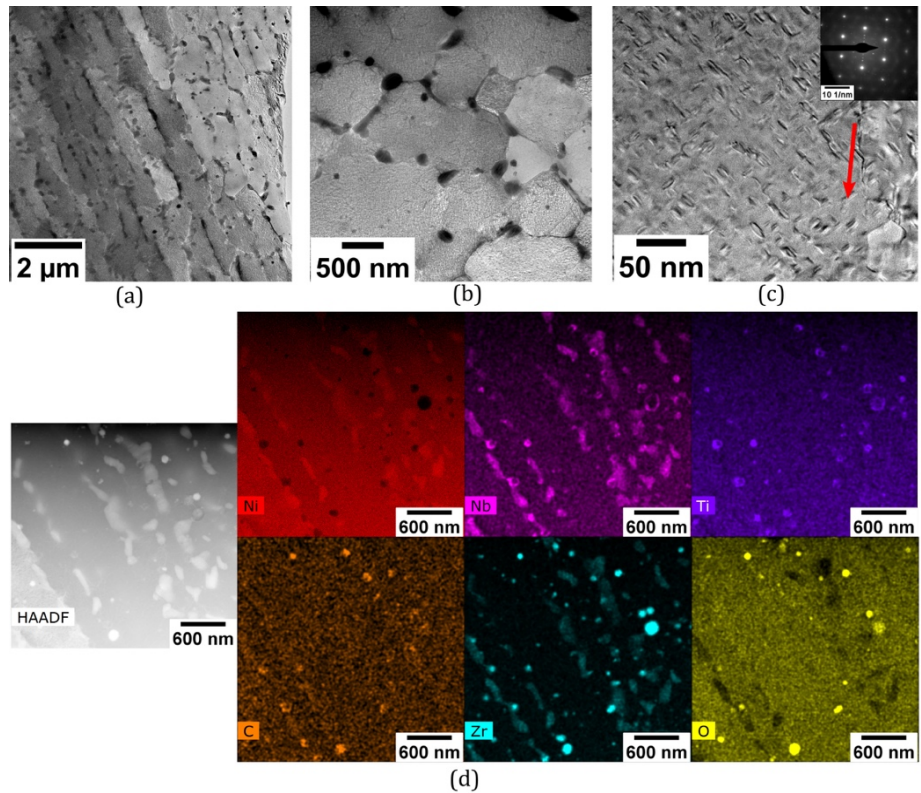

**Fig. 10.** TEM examination of the heat-treated IN718+ZrH$_2$. (a) Cellular structure; (b) globular δ precipitates at the cell boundaries and smaller precipitates within the cells; (c) HAADF-STEM micrograph showing a fine dispersion of disk-shape γ"-Ni$_3$Nb precipitates, which grew parallel to the (001) planes of the Ni FCC lattice. The zone axis of this micrograph is [$\bar{1}$14] and the red arrow corresponds to the direction <220>; (d) HAADF-STEM analysis showing the EDX element maps of the precipitates.

## 3.2. Mechanical properties

### 3.2.1 Effect of heat treatments and Zr particles

The Vickers hardness of the IN718 alloy processed by SLM without and with ZrH$_2$ particles is shown in Table 4 in the as-printed condition and after the heat treatment. The addition of ZrH$_2$ particles increased the hardness by 10% in the as-printed condition but the differences were negligible after the heat treatment.

Table 4. Vickers hardness of IN718 and IN718-ZrH$_2$ processed by SLM.

|  | IN718 (as printed) | IN718-ZrH$_2$ (as-printed) | IN718 (heat-treated) | IN718-ZrH$_2$ (heat-treated) |
|---|---|---|---|---|
| Hardness (HV) | 320±5 | 355±5 | 503±7 | 515±7 |

The engineering tensile stress-strain curves along the building direction of both materials in the as-printed and heat-treated conditions at room temperature are plotted in Figs. 11a and 11b, respectively. The strain was approximated from the cross-head displacement of the mechanical testing machine divided by the gauge length of the sample. The addition of $ZrH_2$ particles led to a marked increase in the yield strength and in the tensile strength in the as-printed condition and also to a marked reduction in the strain-to-failure (Table 5). The strength increased considerably after the heat treatment while the differences in the tensile strength between both alloys disappeared, indicating a change in the mechanisms controlling the strength. Nevertheless, the strain-to-failure of the heat treated alloy with $ZrH_2$ particles was slightly lower.

Table 5. Mechanical properties at room temperature and 550ºC in the building direction of IN718 and IN718-$ZrH_2$ processed by SLM. Strength is expressed in MPa.

|  | IN718 (as printed) | IN718-$ZrH_2$ (as-printed) | IN718 (heat treated) | IN718-$ZrH_2$ (heat treated) |
| --- | --- | --- | --- | --- |
| Strength (23ºC) | 910 ± 6 | 1039 ± 4 | 1371 ± 14 | 1373 ± 4 |
| Strain-to-failure (23ºC) | 0.48 ± 0.01 | 0.19 ± 0.01 | 0.35 ± 0.01 | 0.260± 0.01 |
| Strength (550ºC) | 745 ± 3 | 882 ± 23 | 1088 ± 22 | 1180 ± 30 |
| Strain-to-failure (550ºC) | 0.35 ± 0.04 | 0.24 ± 0.04 | 0.28 ± 0.01 | 0.24 ± 0.04 |

Equivalent engineering tensile stress-strain curves of both materials along the building direction are plotted in Figs. 11c and 11d at high temperature (550ºC) in the as-printed and heat-treated conditions, respectively. Both materials presented good mechanical properties at high temperature (particularly in the heat-treated condition). The presence of Zr particles increased slightly the strength and also reduced slightly the ductility in the as-printed alloy. The mechanical behavior of both materials in the heat treated condition was equivalent (the alloy containing Zr seemed to be a little bit stronger at 550ºC but the differences are within the experimental scatter). Serrations in the stress-strain curves, indicative of dynamic strain ageing (DSA), were observed in the heat-treated condition in both materials when tested at 550ºC (Fig. 11d). DSA was active from the onset of plastic deformation up to failure in IN718 while it was confined to a limited plastic strain range in the material containing Zr particles. Moreover, evidence of DSA close to failure was found in the IN718 alloy in the as-printed condition but not in the alloy containing Zr particles (Fig. 11c)

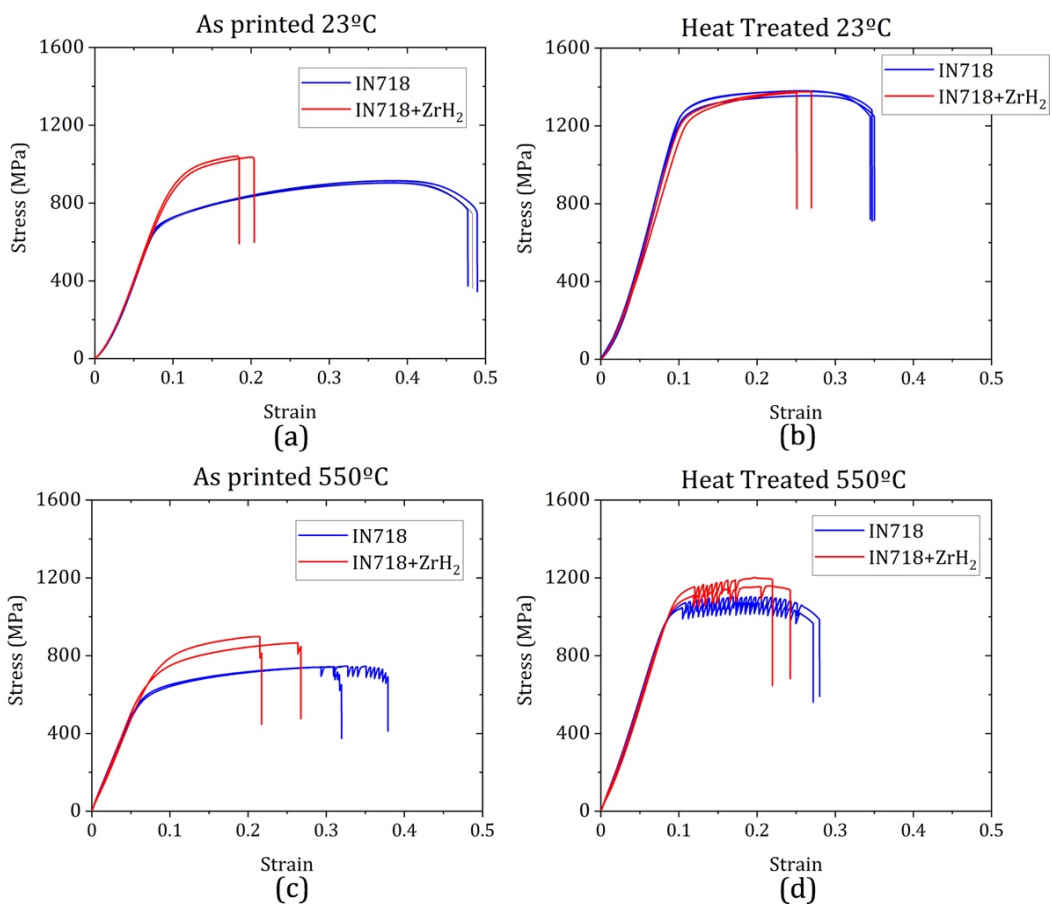

**Fig. 11.** Engineering tensile stress-strain curves along the building direction of the IN718 and IN718 + $ZrH_2$ alloys processed by SLM. (a) As-printed, room temperature (23ºC); (b) heat-treated, room temperature (23ºC); (c) as-printed, high temperature (550ºC); (b) heat-treated, high temperature (550ºC).

3.2.2 Effect of orientation

The effect of orientation on the engineering stress-strain curves of the as-printed and heat-treated IN718 alloy at 23ºC and 500ºC is plotted in Figs. 12a and 12b, respectively. The tests at 550ºC were only carried out in the heat-treated condition. All the mechanical tests showed that the strength of the alloy was higher when deformed perpendicular to the building direction, regardless of the heat-treatment and temperature. Moreover, the difference in strength between both orientations was around 100 MPa in all cases. The strain-to-failure was slightly lower in the specimens deformed perpendicular to the building direction and DSA occurred from the onset of plastic deformation in the specimens tested at 550ºC. All these factors seem to point out that texture was responsible for the observed anisotropy and this hypothesis could be validated from numerical simulations of representative volume elements of the microstructure (see discussion section).

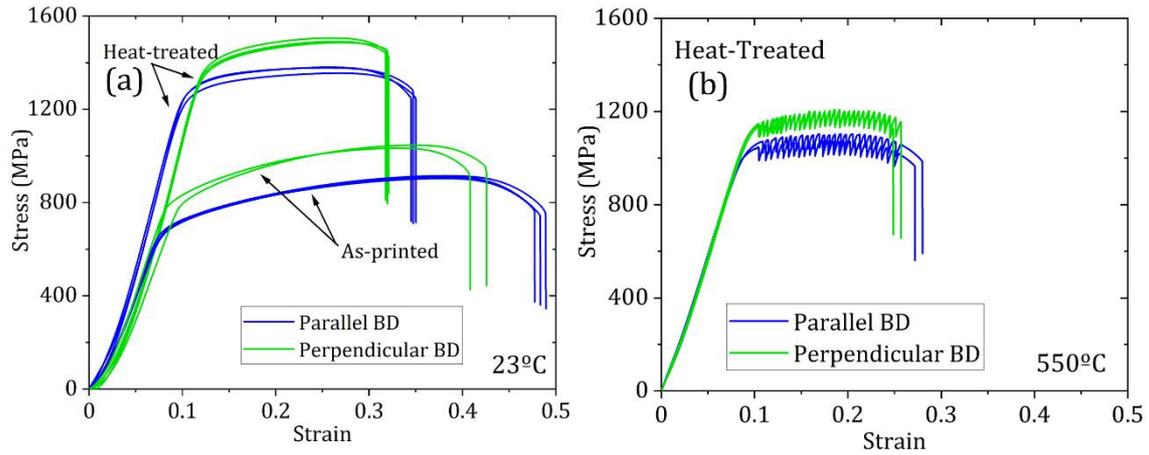

**Fig. 12.** Effect of orientation on the engineering tensile stress-strain curves of the IN718 alloy (parallel and perpendicular to the building direction in blue and green colours). (a) Room temperature (23°C); (b) high temperature (550°C).

# 4. Discussion

4.1 Strengthening mechanisms

Optimization of the SLM processing parameters led to materials with negligible porosities (0.04% and 0.15% for IN718 and IN718+$ZrH_2$, respectively). The slightly higher porosity in the alloy with $ZrH_2$ particles was due to the presence of irregular pores associated with segregation of Zr (Fig. 2). The porosity did not influence the mechanical properties.

Zr particles did not enhance the nucleation rate during SLM processing conditions. Thus, both alloys presented a microstructure formed by elongated grains with the longest axis parallel to the building direction and a strong texture with the (100) plane perpendicular to the building direction. Neither the grain size nor the texture were significantly modified by the addition of $ZrH_2$ particles. However, the IN718+$ZrH_2$ alloy exhibited a higher strength than the IN718 alloy in the as-printed condition, both at 23ºC and at 550ºC, and in the heat-treated condition at 550ºC. The strength of both alloys in the heat-treated condition at ambient temperature was similar (Fig. 11).

The strength of the IN718 alloy in the as-printed condition was determined by the solid solution elements in the $\gamma$ Ni-matrix and by the dimensions of the cells within the grains. The cell boundaries were formed by the segregation of $Ni_3$(Nb, Ti) $\delta$ particles during the rapid solidification conditions that prevail in SLM [10, 28] (Fig. 6). Zr is known to promote the formation of carbides in conventional Ni-based superalloys [29-31] and it is assumed that $ZrH_2$ particles were decomposed due to the high temperatures achieved during laser melting. The Zr was dissolved in the $\gamma$ Ni-matrix and the majority of Zr precipitated upon cooling, leading to a fine dispersion of small carbide particles within the cells (Fig. 7 and 8). These particles also contributed to the strength of the alloy and were mainly responsible for the higher strength of IN718 + $ZrH_2$ alloy in the as-printed condition at both 23ºC and 550ºC. It is also possible that the presence of Ti (that precipitated at the cell boundaries in IN718 but not in IN718 + $ZrH_2$, Figs. 6 and 8) and of Zr in solid solution could contribute to the higher strength of as-printed IN718 + $ZrH_2$.

The strength of both alloys increased dramatically after the heat treatment due to the presence of a fine dispersion of $\gamma$" precipitates (Figs. 9c and 10c). Moreover, the alloys maintained a very high strength at 550ºC. The interaction of dislocations with these precipitates was the dominant strengthening mechanism. Other differences in the microstructure between both alloys (acicular $\delta$ precipitates at grain and cells boundaries in IN718, Fig. 9b, as compared to globular ones in IN718 + ZrH2, Fig. 10b, and the presence of Zr oxide and carbide particles in IN718 + $ZrH_2$, Fig. 10d) did not influence the strength at ambient temperature.

Serrations associated with DSA were found in both heat-treated alloys when deformed at 550ºC (Fig. 11d). Traces of DSA were also found in the as-printed IN718 alloy when tested at the same temperature (Fig. 11c). However, it was obvious that the heat treatment enhanced this mechanism, which could also contribute to the excellent strength retention at 550ºC. DSA has been observed in IN718 manufactured by traditional routes [32-33] as well as by additive manufacturing [34]. In the latter case, DSA was reported in specimens deformed at high temperature (>550ºC) but not at room temperature. It was attributed to the interaction of substitutional atoms (mainly Nb) in solid solution with dislocations [32-34]. The lack of DSA at ambient temperature was associated to the limited mobility of the Nb atoms. The same mechanisms can explain the results in Fig. 11. It is assumed that most Nb was segregated to the grain and cell boundaries after additive manufacturing (Figs. 6 and 8) and, thus, traces of DSA were only found in the as-printed IN718 at the last stages of plastic deformation. The solubilization treatment incorporated the Nb into the $\gamma$ matrix and the subsequent ageing was not able to precipitate all the Nb in the form of $\gamma''$. Some Nb remained in solid solution, leading to the development of DSA in the tensile tests at 550ºC. DSA began at the onset of plastic deformation in the heat-treated IN718 and IN718+ZrH$_2$ (Fig. 11d) but disappeared in the latter while it is maintained until failure in the former. These differences may be attributed to the evolution of the microstructure during deformation, namely deformation-induced precipitation of Nb in the form of carbides that can be promoted by Zr [29-31].

In summary, the higher strength of the IN718 + ZrH$_2$ in the as-printed condition at room and high temperature can be mostly attributed to the formation of carbides, although some beneficial effect of Zr in solid solution cannot be ruled out. The differences in the mechanical behavior become much less noticeable in the heat-treated state due to the dominance of the interaction of dislocations with $\gamma''$ precipitates which mask other, less important effects, derived from the addition of Zr. It is our contention that the occurrence of DSA at high temperatures in both alloys in the heat-treated condition may be attributed to the dissolution of Nb during solubilization.

4.2 Effect of orientation

The consistent differences in the mechanical response of IN718 deformed perpendicular and parallel to the building axis (Fig. 12) could be attributed to the texture induced by SLM. Thus, numerical simulations were carried out to validate this hypothesis. The mechanical response of the polycrystalline IN718 alloy was determined by means of the finite element simulation of a Representative Volume Element (RVE) of the microstructure. The polycrystal was represented by a cubic domain, which was discretized with a regular mesh of 40 x 40 x 40 cubic finite elements (C3D8 elements in Abaqus [35] with 8 nodes and full integration). The RVE was created using Dream3D [36] and contained 100 equiaxed grains whose orientation was selected taking into account the texture of the alloy (Fig. 13).

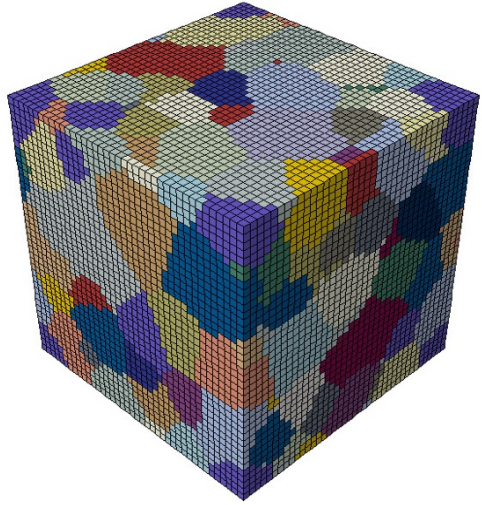

**Fig. 13.** Representative volume element of the microstructure of polycrystalline IN718 containing 100 grains. The grain orientation was obtained from Fig. 4.

The mechanical response of each crystal within the polycrystal was simulated using a phenomenological crystal plasticity model that was successfully used in the past to simulate the mechanical behavior of polycrystalline wrought IN718 [37]. The crystal is assumed to behave as an elasto-viscoplastic solid in which the plastic slip rate in a given slip system follows a power law dependency according to

$$\dot{\gamma}^\alpha = \dot{\gamma}_0 \left| \frac{\tau^\alpha}{\tau_c^\alpha} \right|^{1/m} \mathrm{sgn}(\tau^\alpha) \qquad (1)$$

where $\dot{\gamma}_0$ stands for the reference strain rate, $\tau_c^\alpha$ is the critical resolved shear stress (CRSS) of $\alpha^{th}$ slip system at the reference strain rate and $m$ the rate sensitivity parameter. The evolution of the CRSS of a given slip system $\alpha$, $\tau_c^\alpha$, is expressed as,

$$\dot{\tau}_c^\alpha = \sum_\beta h \left| \dot{\gamma}^\beta \right| \qquad (2)$$

where $\beta$ stands for any slip system. $h$ corresponds to the hardening modulus and its evolution with strain was described according to the Voce hardening model proposed in [38],

$$h(\gamma_a) = h_s + \left[ h_0 - h_s + \frac{h_0 h_s \gamma_a}{\tau_s - \tau_0} \right] \exp^{-h_0 \gamma_a/(\tau_s - \tau_0)} \qquad (3)$$

where $h_0$ is the initial hardening modulus, $\tau_0$ the initial yield shear stress, $\tau_s$ the saturation yields shear stress, $h_s$ the saturation hardening modulus at large strains and $\gamma_a$ is the accumulated shear strain in all slip systems, which is given by

$$\gamma_a = \int_0^t \sum_\alpha |\dot{\gamma}^\alpha| dt \qquad (4)$$

This crystal plasticity model was implemented in a user subroutine UMAT taking into account the 12 {111} <110> slip systems of γ FCC Ni. The elasto-plastic properties of the IN718 single crystals were obtained by Cruzado et al. [37] from micropillar compression experiments in single crystals with different orientation of a wrought IN718 alloy and can be found in Table 6.

**Table 6.** Elastic constants and visco-plastic parameters of the crystal plasticity model of IN718 (from [37])

| $C_{11}$ (GPa) | $C_{12}$ (GPa) | $C_{44}$ (GPa) |
|---|---|---|
| 259.6 | 179 | 109.6 |

| $\dot{\gamma}_0$ (s$^{-1}$) | m | $\tau_0$ (MPa) | $\tau_s$ (MPa) | $h_0$ (GPa) | $h_s$ (GPa) |
|---|---|---|---|---|---|
| 0.001 | 0.017 | 465.5 | 598.5 | 6.0 | 0.3 |

The tensile stress-strain curves in the orientations parallel and perpendicular to building direction were obtained by applying an increasing displacement to the RVE in the corresponding orientation using periodic boundary conditions, following the methodology presented in [39]. The applied strain rate was 2.0 10$^{-3}$ s$^{-1}$ and the engineering stress-strain curves in both orientations are plotted in Fig. 14. Although the absolute values of strength are different from those in Fig. 12 (because the parameters of the crystal plasticity model were not calibrated for IN718 manufactured by SLM) the difference in strength between both orientations (around 100 MPa) is similar to the ones found in Fig. 12. Thus, the strong texture of the SLM IN718 is responsible for the anisotropy in the mechanical behavior.

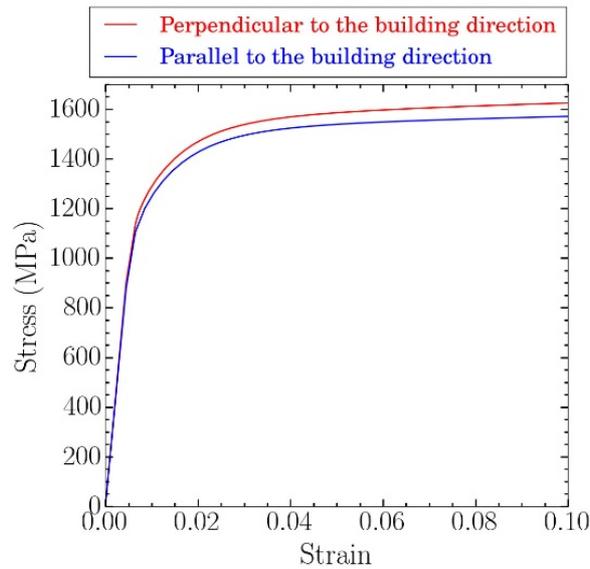

**Fig. 14.** Engineering stress-strain curves as a function of orientation obtained from computational homogenization of a RVE of the microstructure of IN718 polycrystal.

# 5. Conclusions

IN718 powders were mixed with ZrH$_2$ particles in a planetary mill and processed by selective laser melting to ascertain the influence of the particles on the microstructure and mechanical properties of the alloy at room and high temperature. The following conclusions were reached:

- Fully dense alloys (99.96% for IN718 and 99.85% for IN718+ZrH$_2$ particles) were obtained by careful selection of the processing parameters. The slight reduction in density in the alloy with ZrH$_2$ was due to the segregation of Zr.
- The addition of ZrH$_2$ did not modify neither the columnar grain structure not the texture of IN718 manufactured by SLM. The alloys without and with Zr presented a cellular structure after processing with cell walls formed by Nb-rich precipitates. In addition, Zr-rich globular particles with a diameter of ≈ 50 nm were dispersed within the cells in the alloy with Zr. They increased the strength of the as-printed material at both ambient (23ºC) and high temperature (550ºC) as compared with IN718.

- Heat treatments of solubilization followed by aging led to a fine dispersion of $\gamma''$ precipitates in the alloy with and without Zr particles. The strength of both materials was much higher than that of the as-printed alloys and the differences between them were negligible at ambient and elevated temperature. Moreover, dynamic strain ageing was found in both cases when deformed at 550ºC and it was attributed to the interaction of dislocations with Nb atoms in solid solution.
- The strength of the IN718 deformed perpendicular to the building was higher than that along the building direction, regardless of the heat treatment and temperature. Computational homogenization simulations showed that this difference was associated with the strong texture.

# 6. Acknowledgements

This investigation was supported by the European Research Council (ERC) under the European Union's Horizon 2020 research and innovation programme (Advanced Grant VIRMETAL, grant agreement No. 669141). Additional support from the MAT4.0-CM project funded by Madrid region under programme S2018/NMT-4381 is also gratefully acknowledged. CG acknowledges support by the Spanish Ministry of Science, Education and Universities through the Fellowship FPU 18/01328.